\documentclass{sig-alternate}
\usepackage{color}
\usepackage{booktabs}
\usepackage{colortbl}
\newcommand{\argmax}{\operatornamewithlimits{argmax}} 
\newcommand{\indata}{\textbf{P}}

\begin{document}
%
% --- Author Metadata here ---
% \conferenceinfo{ACM RecSys}{'09 New York, NY USA}
%\CopyrightYear{2009} % Allows default copyright year (200X) to be over-ridden - IF NEED BE.
%\crdata{0-12345-67-8/90/01}  % Allows default copyright data (0-89791-88-6/97/05) to be over-ridden - IF NEED BE.
% --- End of Author Metadata ---

\title{Recommender Systems \\
for the Conference Paper Assignment Problem}
%\subtitle{[Extended Abstract]
%\titlenote{A full version of this paper is available as
%\textit{Author's Guide to Preparing ACM SIG Proceedings Using
%\LaTeX$2_\epsilon$\ and BibTeX} at
%\texttt{www.acm.org/eaddress.htm}}}
%
% You need the command \numberofauthors to handle the 'placement
% and alignment' of the authors beneath the title.
%
% For aesthetic reasons, we recommend 'three authors at a time'
% i.e. three 'name/affiliation blocks' be placed beneath the title.
%
% NOTE: You are NOT restricted in how many 'rows' of
% "name/affiliations" may appear. We just ask that you restrict
% the number of 'columns' to three.
%
% Because of the available 'opening page real-estate'
% we ask you to refrain from putting more than six authors
% (two rows with three columns) beneath the article title.
% More than six makes the first-page appear very cluttered indeed.
%
% Use the \alignauthor commands to handle the names
% and affiliations for an 'aesthetic maximum' of six authors.
% Add names, affiliations, addresses for
% the seventh etc. author(s) as the argument for the
% \additionalauthors command.
% These 'additional authors' will be output/set for you
% without further effort on your part as the last section in
% the body of your article BEFORE References or any Appendices.

\numberofauthors{1} %  in this sample file, there are a *total*
% of EIGHT authors. SIX appear on the 'first-page' (for formatting
% reasons) and the remaining two appear in the \additionalauthors section.
%
\author{Don Conry$^\dagger$, Yehuda Koren$^\$$, and Naren Ramakrishnan$^\dagger$\\
$^\dagger$Department of Computer Science, Virginia Tech, VA 24061, USA\\
$^\$$Yahoo! Research, Israel}

% There's nothing stopping you putting the seventh, eighth, etc.
% author on the opening page (as the 'third row') but we ask,
% for aesthetic reasons that you place these 'additional authors'
% in the \additional authors block, viz.
% \additionalauthors{Additional authors: John Smith (The Th{\o}rv{\"a}ld Group,
% email: {\texttt{jsmith@affiliation.org}}) and Julius P.~Kumquat
% (The Kumquat Consortium, email: {\texttt{jpkumquat@consortium.net}}).}
% \date{30 July 1999}
% Just remember to make sure that the TOTAL number of authors
% is the number that will appear on the first page PLUS the
% number that will appear in the \additionalauthors section.

\maketitle
\begin{abstract}
Conference paper assignment, i.e., the task of assigning paper submissions to reviewers, presents multi-faceted issues for recommender systems research. Besides the traditional goal of predicting `who likes what?', a conference management system must take into account aspects such as: reviewer capacity constraints, adequate numbers of reviews for papers, expertise modeling, conflicts of interest, and an overall distribution of assignments that balances reviewer preferences with conference objectives. Among these, issues of modeling preferences and tastes in reviewing have traditionally been studied separately from the optimization of paper-reviewer assignment. In this paper, we present an integrated study of both these aspects. First, due to the paucity of data per reviewer or per paper (relative to other recommender systems applications) we show how we can integrate multiple sources of information to learn paper-reviewer preference models. Second, our models are evaluated not just in terms of prediction accuracy but in terms of the end-assignment quality. Using a linear programming-based assignment optimization formulation, we show how our approach better explores the space of unsupplied assignments to maximize the overall affinities of papers assigned to reviewers. We demonstrate our results on real reviewer preference data from the IEEE ICDM 2007 conference.
\end{abstract}

\vspace{-0.1in}
% A category with the (minimum) three required fields
%A category including the fourth, optional field follows...
%\terms{Algorithms, Human Factors}

\vspace{-0.1in}

\section{Introduction}
%Recommender systems have traditionally been studied in e-commerce domains
%such as books, food, 
%music, and movies, but are applicable in practically any
%domain where the goal is to model user preferences and achieve
%objectives of bringing people and artifacts together. Here we study
%the use of recommender systems for the
%conference paper assignment problem (CPAP).
%Just as e-commerce product recommenders
%that model partial data about people and the
%products they like, the goal in CPAP is to model data about people and
%the papers they would like to review.
%
Modern conferences, especially
in areas such as data mining/machine learning (KDD; ICDM; ICML; NIPS) and
databases\hskip0ex /web (VLDB; SIGMOD; WWW), are beset with 
excessively high numbers of paper submissions. Assigning these papers to
appropriate reviewers in the program committee (which can constitute a few hundred
members) is a herculean task and hence motivates the use of recommender systems.

%This motivates the study of recommender systems
%for the conference paper assignment problem (CPAP).
Besides the
traditional goal of
predicting `who likes what?', a conference management system must take into account aspects such as: reviewer capacity constraints, adequate numbers of reviews for papers, expertise modeling, conflicts of interest, and an overall distribution of assignments that balances reviewer preferences with conference 
objectives. 
Among these, issues of modeling preferences, expertise, 
and tastes in reviewing have traditionally been studied separately from the optimization of paper-reviewer assignment. The former has been the subject of
much academic research (see Section~\ref{sec:litrev1}) while the latter is emphasized by commercial 
software, such as 
EasyChair, CyberChair, and Microsoft's CMS, which aim to automate
the management of the conference reviewing process.  

We investigate the conference paper assignment problem
(CPAP) through the lens of recommender systems research.
There are three key differences from traditional recommender systems
research and the CPAP problem. 
First, in a traditional recommender, recommendations that meet the needs of
one user do not affect the satisfaction of other users. In CPAP, on the other
hand, multiple users (reviewers) are bidding to review the same papers and hence
there is the possibility of one user's recommendations (assignments) 
affecting the satisfaction
levels (negatively) of other users. Hence the design of reviewer
preference models
must be posed and studied in an overall optimization framework.

%These problems can 
%be studied in a game theoretic sense but we take a conventional approach 
%here toward building preference models to seed the assignment process.
Second, in a conventional recommender,
the goal is often to recommend {\it new} entities that are likely to be
of interest, whereas in CPAP, the goal is to ensure that reviewers
are predominantly assigned their (most) preferred papers. Nevertheless,
preference modeling is still crucial because it gives 
the assignment algorithm some degree of latitude in aiming 
to satisfy multiple users.

Finally, recommender systems are used to working with sparse data but
the amount of `signal' available to model preferences in the CPAP
domain is exceedingly small;
hence we must integrate multiple sources of information 
to build strong preference models.

In this paper, we present the first integrated study of both modeling
reviewing preferences and optimizing assignments for conference
management. Our key contributions can be summarized as follows.
\begin{enumerate}
\item 
Due to the paucity of data per reviewer or per paper (relative to other recommender systems applications) we show how we can integrate 
information about publication subject categories, contents of paper
abstracts, and co-authorship information to learn improved paper-reviewer
preference models.
\item We evaluate our models not just 
in terms of prediction accuracy but in terms of the end-assignment quality. Using a linear programming-based assignment optimization formulation, we show how our approach better explores the space of unsupplied assignments to maximize the overall affinities of papers assigned to reviewers. 
\item We demonstrate the effectiveness of our approach on actual reviewing
preference data in the context of a real life conference, 
namely the IEEE ICDM'07 conference~\cite{icdm2007}.
\end{enumerate}

\section{Related Research}
Any conference management system must contend with two main issues: 
how to model affinities or preferences between papers and reviewers, and 
how to use these affinities to make and/or optimize assignments. For the former issue,
many conferences have an explicit `bidding' phase and use data collected
in this phase as the affinity matrix. While many conferences use these
bids as-is, we will
demonstrate how they can be
used as the starting point to build improved
preference models. Approaches to solve the latter
issue have traditionally
been considered orthogonal to the problem
of preference modeling but, as we demonstrate later, 
better preference modeling leads to improvements in this phase as well.

\subsection{Modeling Affinities, Preferences, and Expertise} \label{sec:litrev1}
The sparsity of reviewer-paper bidding data has led some researchers,
e.g., Rigaux~\cite{rigaux2004}, to explore the use of collaborative
filtering techniques~\cite{tapestry,grouplens} to `grow' the given bids.
The underlying assumption is that reviewers who bid similarly on 
a number of the same papers are likely to have similar preferences for
other papers. Basu et al~\cite{basu2001} use the relational WHIRL system
to integrate similarity scores from disparate data sources to identify most
relevant (paper,reviewer) combinations. They do not, however, 
%\narenc{Don, does the basu/popescul paper deal with conference paper assignment?
%If not, we should remove it. Also, it looks like they are modeling
%similarity scores? Not paper-reviewer scores? If so, come out and say it.}
%\donc{Yes (conf. assignment) and the similarity scores are used \emph{as} paper-reviewer scores (perhaps indirectly?). Removed citation to 2nd Basu paper below, however.}
%Basu et al~\cite{basu2001} use the relational WHIRL system to multiply similarity scores from disparate data sources into a single aggregate similarity score. WHIRL uses methods from the field of Information Retrieval; it constructs a vector space (TF-IDF) similarity model to compare papers or reviewers. The system is then 'queried' for the top \textbf{n} similar items for a given paper or reviewer. This system is very similar to search in IR, and thus results can be ranked by 'precision'. The authors do not 
attempt to satisfy per-paper or per-reviewer constraints, and the contributions of different sources are considered equivalent to each other.
Popescul et al~\cite{popescul2001} present a way to combine content-based and collaborative recommendations using a three-way aspect model. 
%The authors use conditional probabilities and a statistical technique known as \textit{expectation maximization} to discover optimal weights for the different data sources. It is suggested that especially in conditions where primary data is sparse, secondary data improves the quality of recommendations by simulating increased data density. This is similar in direction to the approach presented below, but uses a probabilistic model instead.
%
The GRAPE system~\cite{GRAPE} prefers topical information over supplied reviewer bids 
or preferences, but does use the preferences as a secondary means of modeling. The
rationale is the view that
%; however, preferences \textit{are} used as a secondary means of 
%assignment (i.e. are given less weight in the assignment process). The reasoning behind this decision is the assumption that the 
topical data more accurately predicts the degree of expertise present for 
a reviewer-paper match. Since the distribution of reviewers and papers over topics is unpredictable (sometimes leaving too many or too few reviewers for a given cluster of papers), the preference information is used for tuning or smoothing out the wrinkles in the topic-based assignments.

%More sophisticated forms of expertise modeling have been explored in,
%e.g.,~\cite{mccallum2007}. 
%One view of conference review assignments is that papers should be assigned to reviewers with a certain degree of expertise in the specific field or topic of the paper. This view leads to topic-based approaches that use additional information to improve paper assignments; namely, the main topic or topics of each paper, as well as (in some cases) the area or areas of expertise for each reviewer. By using this information, reviewer assignments can be tailored to suit each paper's topic, ensuring a degree of expertise is present. The resultant ranking of each reviewer based on topical knowledge with respect to a given paper has been called \textit{expert-finding} and \textit{expertise modeling}~\cite{mccallum2007}.
%
A problem faced by most expertise modeling approaches is
identifying which topics are covered in papers. Early efforts in this area focused mainly on paper abstracts, and topical expertise was determined through common information retrieval methods involving keywords. For example, Dumais and Nelson~\cite{dumais1992} match papers to reviewers using Latent Semantic Indexing (LSI) trained on reviewer-supplied abstracts. 
%In Basu et al \cite{basu1999}, abstracts from papers written by potential reviewers are extracted from the web via search, and a vector space model is constructed for matching. \donc{removed for space concerns in bibliography, also see alternate Basu paper cited above with similar content.}
Yarowsky and Florian~\cite{yarowsky1999} extended this idea by using a similar vector space model with a naive Bayes classifier on work previously published by each reviewer.

More recently, Wei \& Croft~\cite{weicroft2006} describe a topic-based model using a language model with Dirichlet smoothing.
An excellent example of topic-based models is the Author-Persona-Topic (APT) model by Mimno \& McCallum~\cite{mccallum2007}. The APT model contains a number of features designed to better capture the reality of the relationship between conference reviewers and papers. The idea is that an author may study and write about several distinct topics; by clustering papers from each of these topics into a separate persona for an author, the author's ranking for a given topic need not be diluted by his or her writings on a different topic.
% Again, the system was trained using abstracts of the potential reviewers, and the APT model was found to perform better than other language-based and topic-based models.

\subsection{Optimizing Assignments}
\label{optimizations}
Given preference data, either explicitly gathered or computationally modeled,
the actual task of making assignments can be viewed as bipartite
matching. 
%\narenc{Define the bipartite matching problem here - its
%vanilla version, then talk about other versions, such as weighted matching,
%matching with multiple edges, stable matching, etc.}
%A popular approach is to consider CPAP as a bipartite matching problem. This approach models reviewers and papers as vertices of a graph connected via weighted edges; the objective is to connect the vertices of one type to the other, while maximizing the sum of the edge weights. 
The classical approach to bipartite matching is given by the Hungarian Algorithm described by Kuhn~\cite{kuhn1955}; it provides a solution for the simplest cases of this family of problems (applicable when the number of reviewers equals the number of papers). Various refinements have been made to this algorithm over the years, such as one by Hopcroft and Karp~\cite{karp1973}. A number of contemporary assignment systems take this approach, including GRAPE~\cite{GRAPE}.
%\donc{perhaps this diagram helps? otherwise, could reclaim this space.}
%\newcommand{\edge}[1]{\ar@{.}[#1]}
%\newcommand{\chosen}[1]{\ar@{->}[#1]}
%\newcommand{\chosenup}[1]{\ar@{->}@/^12pt/[#1]}
%\newcommand{\chosendown}[1]{\ar@{->}@/_12pt/[#1]}
%\newcommand{\curvu}[1]{\ar@/^12pt/@{.}[#1]}
%\newcommand{\curvd}[1]{\ar@/_12pt/@{.}[#1]}
%\newcommand{\node}{*+[o][F-]{ }}
%\newcommand{\nodebox}{*+[F*]{ }}
%\begin{center}
%\begin{figure}[!ht]
%\label{figure2}
%\centerline{
%\xymatrix{
%& \textbf{Alice} & \node{} \chosen{rrrr}|{\textbf{1}} \curvd{rrrrdd}|2 \curvd{rrrrdddd}|2 & & & & \nodebox{} & \textit{1st paper} \\\
%& \\
%& \textbf{Bob} & \node{} \edge{rrrr}|>>>>>>>>>>>>>>>>>>4 \chosendown{rrrrdd}|{\textbf{2}} \curvu{rrrruu}|5 & & & & \nodebox{} & \textit{2nd paper} \\
%& \\
%& \textbf{Charlie} & \node{} \edge{rrrr}|-3 \chosenup{rrrruu}|{\textbf{1}} \curvu{rrrruuuu}|3 & & & & \nodebox{} & \textit{3rd paper} \\
%}
%}
%\caption{A simple bipartite matching problem, as discussed in \cite{rigaux2004}.}
%\end{figure}
%\end{center}
%
For practical reasons, it is useful to restrict the number of reviews per reviewer and per paper; a constraint based linear program, e.g., work by Taylor~\cite{taylor2006}, 
is a natural approach. 
%Taylor~\cite{taylor2006} proposes to add an additional \textit{affinity} constraint to represent the expertise factor. The set of assignments can then be quantitatively judged based on the global affinity attained. Taylor uses area chairs to rank authors based on personal knowledge of each reviewer's expertise. Another possibility would be to ask the reviewers themselves to indicate their preferences for reviewing (or not reviewing) each paper. These affinity values can be compiled into a matrix of all reviewer-paper combinations, and used to solve the original constraint-satisfaction problem. The affinity values effectively rank reviewers for each paper, similar to the effect of the other models above.
%
%\narenc{lead into network flow gently.}

Another approach to CPAP uses reasoning from the much more general \textit{minimal cost network flow} problems studied in dynamics and operations research. Many such related problems (known collectively as extended Generalized Assignment Problems~\cite{benferhat2001} or GAP) of assigning a limited number of resources to certain tasks exist in diverse fields. In the network flow diagram of this general assignment problem, resources (in our case, reviewers) are represented by source nodes with a certain supply (number of reviews allowed per reviewer), while tasks (each paper to be reviewed) are sink nodes with a demand (the number of times each paper must be reviewed). For
specific approaches, see~\cite{goldsmith2007,hartvigsen1999}.

\begin{figure*}[!ht]
\centering
\includegraphics[width=6.5in]{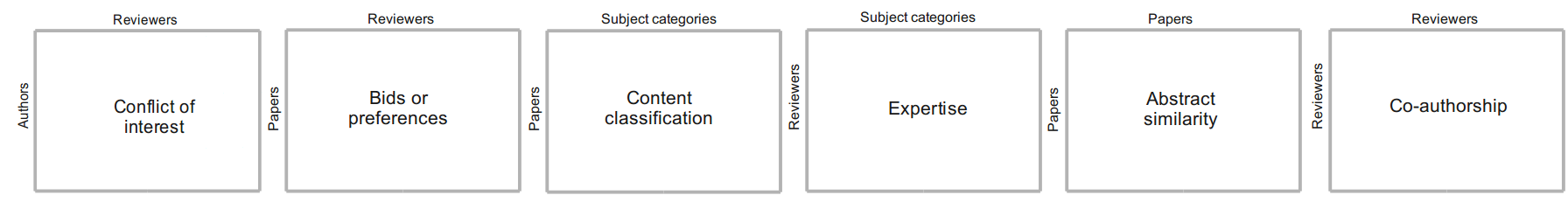}
\caption{Data used in this paper for building paper-reviewer preference models.}
\label{fig:data}
\end{figure*}

\section{Models of Review Preferences} \label{sec:Prefmodels}
We adapt recommendation techniques for predicting unknown reviewer-paper preferences. Naturally, reviewers assume the role of ``users'' in traditional recommender systems, while papers take the role reserved to ``products''.  Our goal is to exploit a variety
of available information (see Fig.~\ref{fig:data})
in order to get better estimates of those unknown preferences. This, in turn, will allow the assignment algorithm to find better matches between reviewers and papers. First, we introduce some essential conventions.

\subsection{Notation and Dataset Description}

We are given {\em ratings} (henceforth, interchangeable with {\em preferences}) about $m$ reviewers and $n$ papers. We reserve special indexing letters for distinguishing reviewers from papers: for reviewers $u,v$, and for papers $i,j$. A rating $r_{ui}$ indicates the preference by reviewer $u$ of paper $i$, where high values mean stronger preferences. Usually the vast majority of ratings are unknown. 

As a concrete example, the dataset utilized in this paper comes from the
Seventh IEEE International Conference on Data Mining (ICDM'07) held in Omaha, NE, USA
(utilized here with permission).
The originally supplied matrix is sparse: 529 papers, 203 reviewers, and only
6267 bids. This means that a reviewer rates about 31 papers on average, while a paper recieves less than 12 ratings on average.
Each rating reflects a bid a reviewer put on a paper, with numerical values, 
between 1 and 4, indicating preferences as follows: 4= ``High'', 3=``OK'', 
2=``Low'' and 1=``No''.

%The results in this paper are based on supporting data obtained from the International Conference on Data Mining, for the year 2007 (ICDM'07~\cite{icdm2007}). This data is given in matrix form; it includes reviewers 'bid' or preference data, as well as topical expertise of reviewers, and content classification for papers. The preference data consists of reviewer responses indicating their willingness (or lack of it) to review each paper. Since ICDM includes a relatively large number of submitted papers, most reviewers bid on a small subset of papers (and some do not return bids at all). As a result, the preference matrix is sparse; out of 529 papers and 203 reviewers (107,387 potential combinations), 6,267 bids were entered (less than 6 percent).
%
%in our dataset which is based on data collected from a recent data mining conference, the number reviewers is 203, the number of papers is 529. Overall we are given 6267 reviewer-paper preferences, which mean that a reviewer rates about 31 papers on average. 
%

We distinguish predicted ratings from known ones, by using the notation $\hat{r}_{ui}$ for the predicted value of $r_{ui}$.  To evaluate the models we split the dataset into a train set, which contains about 90\% of the preferences (randomly chosen), and a test set, which contains the rest preferences. Consequently, our models learn the train set and assign values to $\hat{r}_{ui}$ for all $(u,i)$-pairs in the test set.
Results from these runs are averaged over 100 iterations of training-test data
splits.

The quality of the results on a specific test set ($TestSet$) is measured by their root mean squared error (RMSE):\\
$
\sqrt{{\sum_{(u,i) \in TestSet}(r_{ui}-\hat{r}_{ui})^2}/{|TestSet|}} 
$.
The overall accuracy of the model is taken as the mean RMSE over the
100 randomly generated test sets. The reason for using such a randomization is the small size of our dataset, which makes each individual test set relatively small. 

We hasten to add that we do not advocate the myopic view of RMSE~\cite{rmse-bad} as 
the primary criterion for recommender systems evaluation. We use it in this
section primarily due to its convenience for constructing direct optimizers.
In the next section we will evaluate performance according to criteria more natural to the paper assignment problem. We also note that small improvements in overall
RMSE will typically translate into substantial improvements in bottom-line performance for predicting paper-reviewer preferences.

In the following, we gradually expand the prediction model, by introducing into it a growing set of features.

\subsection{Baseline model}

Much of the variability in the data is explained by global effects, which can be reviewer- or paper-specific. It is important to capture this variability by a separate component, thus letting the more involved models deal only with genuine reviewer-paper interactions. We model these global effects through:
\begin{equation} \label{model1}
 \hat{r}_{ui} = \mu + b_u + b_i
\end{equation}
The constant $\mu$ indicates a global bias in the data, which is taken to be the overall mean rating. The parameter $b_u$ captures reviewer-specific bias, accounting for the fact that different reviewers use different rating scales. Finally, the paper bias, $b_i$, accounts for the fact that certain papers tend to attract higher (or, lower) bids than others.

We learn optimal values for $b_u$ ($u=1,\dots,m$) and $b_i$ ($i=1,\dots,n$), by minimizing the associated squared error function (or, equivalently, the train RMSE):
$$
\min_{b_*} \sum_{(u,i)  \in TrainSet} (r_{ui}-\mu - b_u - b_i)^2 +\lambda_1  b_u^2  +\lambda_2 b_i^2
$$
The regularizing term, i.e., $\lambda_1  b_u^2  +\lambda_2 b_i^2$,
avoids overfitting by penalizing the magnitudes of the parameters. We set the values of the constants   $\lambda_1$ and $\lambda_2$ by cross validation. Learning is done by stochastic gradient descent (alternatively, any least squares solver could be used here).
The resulting average test RMSE is {\bf 0.6286}.

A separate analysis of each of the two biases shows reviewer effect ($\mu + b_u$, with RMSE {\bf 0.6336}) to be much more significant than paper bias ($\mu + b_i$, RMSE {\bf 1.2943}) in reducing the error. This indicates a tendency of reviewers to concentrate all  ratings near their mean ratings, which is supported by examination of the data.

While the baseline model could explain much of the data variability, as evident by its relatively low associated RMSE, it is useless for making actual assignments. After all, it gives all reviewers exactly the same order of paper preferences. Thus, we are really after the remaining unexplained variability, where reviewer-specific preferences are getting expressed. Uncovering these preferences is the subject of the next subsections.

\subsection{A factor model}

Latent factor models comprise a common approach to collaborative filtering with the goal to uncover latent features that explain observed ratings; examples include pLSA \cite{plsa_cf}, neural networks \cite{rbm}, and Latent Dirichlet Allocation~\cite{lda}. We will focus on models that are induced by factorization of the reviewer-paper ratings matrix, which recently have gained popularity \cite{canny,koren2008,paterek,toronto,gravity-sigkddexplore}, thanks to their attractive accuracy and scalability.  

The premise of such models is that both reviewers and papers can be characterized as vectors in a common $f$-D space. The interaction between reviewers and papers is modeled by inner products in that space. Together, with the non-interaction signal covered in the previous subsection, a rating is predicted by the rule:
\begin{equation} \label{model2}
 \hat{r}_{ui} = \mu + b_u + b_i + p_u^Tq_i
\end{equation}
Here, $p_u \in \mathbb{R}^f$ and $q_i \in \mathbb{R}^f$ are the factor vectors of reviewer $u$ and paper $i$, respectively. These are learnt by minimizing the associated squared error function, using stochastic gradient descent.
The resulting average test RMSE is slowly decreasing when increasing the dimensionality of the latent factor space. E.g., for $f=50$ it is {\bf 0.6240}, and for $f=100$ it is {\bf 0.6234}. Henceforth, we use $f=100$.

\subsection{Subject categories}

While latent factor models automatically infer suitable categories, much can be learnt by known categories attributed to both papers and reviewers. In a typical conference submission process, authors are requested to denote primary and secondary categories appropriate for their papers. Likewise, reviewers are asked to indicate their interest along the same categories. It would be desirable to match reviewers with papers lying within their area of expertise.

More specifically, for our dataset, which contains a number of predefined categories judged relevant for ICDM'07 (see Table~\ref{table:categories}), the entered matching between paper $i$ and category $c$ is denoted by:
$$
\sigma_{ic} = \left\{ 
 \begin{array}{cc}
 1 &  c \in {\rm primary}(i) \\
\frac12 & c \in {\rm secondary}(i) \\
 0 & {\rm otherwise} 
 \end{array}\right.
 $$
The value assignment (1 for ``primary'', 0.5 for ``secondary'') is derived by cross validation and is quite intuitive. Similarly, we use the following for matching reviewers with their desired categories:
$$
\theta_{uc} = \left\{ 
 \begin{array}{cc}
 1 &  c \in {\rm interest}(u) \\
-\frac12 & c \in {\rm conflict}(u) \\
 0 & {\rm otherwise} 
 \end{array}\right.
 $$
Notice that in our dataset, reviewers could enter negative interest in certain categories, with which they have a ``conflict of interest''.

This leads to a model, which measures the interaction between reviewers and papers based on the association of the respective entered categories, leading to:
\begin{equation} \label{model3}
 \hat{r}_{ui} = \mu + b_u + b_i + \sum_c \sigma_{ic} \theta_{uc} w_c
\end{equation}

The weights $w_c$ indicate the significance of each category in linking a reviewer to a paper. Those are learnt automatically from the data by minimizing the squared error on the train set. It is plausible that, e.g., a mutual interest in some category A, will strongly link a reviewer to a paper, while a mutual interest in another category B is less influential on papers choice. For a concrete example, refer to Table~\ref{table:categories}, which shows the categories in our dataset sorted by their respective $w_c$ values. We observe differences of orders of magnitude in the ability of different categories to correctly predict associations of reviewers to papers. Note in 
particular that there is no obvious monotonic relationship between the weight imputed
to categories and the number of papers/reviewers associated with the category.

The resulting average test RMSE of the model is: {\bf 0.6243}. This can be improved by integrating with the latent factor model, yielding:
\begin{equation} \label{model4}
 \hat{r}_{ui} = \mu + b_u + b_i+ p_u^Tq_i + \sum_c \sigma_{ic} \theta_{uc} w_c
\end{equation}
The RMSE here is {\bf 0.6197}.

\begin{table*}[tb]
\caption{Subject categories used for associating reviewers and papers. Categories are ranked by their weights, which indicate the ability of each category to
match papers to appropriate reviewers, as learnt by our model. For comparison
the number of papers (assigned to the topic) and reviewers (claiming expertise
in the topic) are also shown.}
\label{table:categories}
\vspace{0.1in}
\centering
{\scriptsize
\begin{tabular}{l|c|c|c}
{\bf Category} & {\bf Weight} & {\bf \# reviewers} & {\bf \# papers}\\
& & & primary (secondary) \\
\hline
Healthcare, epidemic modeling, and clinical research & 0.395121 & 31 & 7 (7) \\
Security, privacy, and data integrity & 0.334821 & 23 & 12 (6) \\
Handling imbalanced data & 0.284398 & 24 & 6 (10) \\
Data mining in electronic commerce, such as recommendation,& 0.260062 & 39 & 16 (19) \\ ~~~sponsored web search, advertising, and marketing tasks & && \\
Mining textual and unstructured & 0.245319 & 66 & 38 (30) \\
Intrusion detection, fraud prevention, and surveillance	& 0.23251 & 28 & 7 (12) \\
Statistical foundations for robust and scalable data mining & 0.228847 & 23 & 9 (16) \\
Quality assessment, interestingness analysis, and post-processing & 0.21166 & 30 & 11 (12) \\
Mining in networked settings: web, social and computer networks,&0.206318 & 62 & 44 (29) \\ ~~~and online communities &&&  \\
Mining high speed data streams& 0.172367 & 40 & 18 (8) \\
Human-machine interaction and visual data mining & 0.168258 & 23 & 7 (9) \\
Telecommunications, network and systems management & 0.152845 & 11 & 2 (3) \\
Computational finance, online trading, and analysis of markets & 0.11785 & 18 & 5 (6) \\
Bioinformatics, computational chemistry, geoinformatics, & 0.108648 & 51 & 14 (26) \\ ~~~and other science engineering disciplines&&&\\
Mining sequences and sequential data & 0.102578 & 57 & 20 (19) \\
Automating the mining process and other process related issues & 0.098819 & 10 & 6 (8) \\
Novel data mining algorithms in traditional areas (such as classification, & 0.089248 & 91 & 147 (71) \\~~~regression, clustering, probabilistic modeling, and association analysis) &&&  \\
Mining spatial and temporal datasets & 0.081676 & 45 & 22 (16) \\
Customer relationship management & 0.081414 & 21 & 0 (3) \\
Mining sensor data & 0.05508 & 40 & 8 (12) \\
Dealing with cost sensitive data and loss models & 0.03453 & 12 & 4 (4) \\
Data pre-processing, data reduction, feature selection, and feature & 0.012069 & 46 & 33 (43) \\ ~~~transformation &&& \\
High performance implementations of data mining algorithms & 0.008198 & 38 & 13 (24) \\
Algorithms for new, structured, data types, such as arising in & 0.006015 & 60 & 21 (25) \\
~~~chemistry, biology, environment, and other scientific domains &&& \\
Distributed data mining and mining multi-agent data & 0.000255 & 29 & 4 (8) \\
Developing a unifying theory of data mining & 0 & 36 & 4 (7) \\
\hline
\end{tabular}}
%  \narenc{Give
% two extra columns. For each subject, list the number of reviewers who
% declared that as their area of expertise, and also the number of papers
% that marked it as primary/secondary area. e.g., say 6(2) meaning 6 papers
% chose that as primary and 2 as secondary. This is so that we can see whether
% higher weights are simply given to those with higher numbers, or something
% more funky is going on.}}
\vspace{-3mm}
\end{table*}

\subsection{Paper-paper similarities}

We inject paper-paper similarities into our models in a way reminiscent of item-item recommenders \cite{item-item}. The building blocks here are similarity values $s_{ij}$, which measure the similarity of paper $i$ and paper $j$. The similarities could be derived from the ratings data, but those are already covered by the latent factor model. Rather, we derive the similarity of two papers by computing the cosine of their abstracts. Usually we work with the square of the cosine, which better contrasts the higher similarities against the lower ones.

This leads to a model where a reviewer's preference for a paper is derived from his preferences to similar papers, through a weighted average, as follows:
\begin{equation} \label{model5}
 \hat{r}_{ui} = \mu + b_u + b_i + \gamma \frac{\sum_{j \in {\rm R}(u)}s_{ij} r_{uj}}{\alpha+\sum_{j \in {\rm R}(u)}s_{ij}}
\end{equation}

Here, the set ${\rm R}(u)$ contains all papers on which $u$ bid. The constant $\alpha$ is for regularization: it is penalizing cases where the weighted average has very low support, that is $\sum_{j \in {\rm R}(u)}s_{ij}$ is very small (e.g., no similar paper was rated by $u$). In our dataset it was determined by cross validation to be 0.001. The parameter $\gamma$ sets the overall weight of the paper-paper component. It is learnt as part of the optimization process (cross-validation could have been used as well). Its final value is closely 0.7. Overall, the resulting average test RMSE of this model is {\bf 0.6109}, which is better than what other models could achieve so far.

As usual, we combine the paper-paper similarities into our overall scheme, which further drops RMSE to: {\bf 0.6038}, through the following model:
\begin{equation} \label{model6}
 \hat{r}_{ui} = \mu + b_u + b_i + p_u^Tq_i + \sum_c \sigma_{ic} \theta_{uc} w_c + \gamma \frac{\sum_{j \in {\rm R}(u)}s_{ij} r_{uj}}{\alpha+\sum_{j \in {\rm R}(u)}s_{ij}}
\end{equation}

\subsection{Reviewer-reviewer similarities}

In analogy to paper-paper similarities, one can also use reviewer-reviewer similarities, in order to borrow preferences between like minded reviewers. This is reminiscent of classic user-user collaborative filtering. Once again, we do not want to derive user-user relations directly from their preferences, as the signal from there is already incorporated into the latent factor model. Instead, we resort to an additional data source for deriving those similarities. Here, one can use the publication histories of the reviewers. To model reviewer-reviewer similarities, we utilize the number of
commonly co-authored papers as reported in DBLP,
denoted by $s_{uv}$. (More sophisticated choices are of course
 open for future exploration.) In parallel to the paper-paper model, a preference can be predicted by following the rule:

\begin{equation} \label{model7}
 \hat{r}_{ui} = \mu + b_u + b_i + \phi \frac{\sum_{v \in {\rm R}(i)}s_{uv} r_{vi}}{\beta+\sum_{v \in {\rm R}(i)}s_{uv}}
\end{equation}

Here, the set ${\rm R}(i)$ contains all reviewers that rated $i$. The regularizing constant $\beta$ is penalizing cases where the weighted average has very low support, that is, $\sum_{v \in {\rm R}(i)}s_{uv}$ is very small (e.g., no similar reviewer has rated $i$). It was determined by cross validation to be 0.001. The parameter $\phi$ sets the overall weight of the reviewer-reviewer component. It is learnt as part of the optimization process, with final value close to 0.06 for our dataset. (Notice that $\phi$ is much smaller than the analogous weight of the paper-paper component, $\gamma=0.7$.) Overall, the resulting RMSE of this model is {\bf 0.6262}, thus offering less accuracy than its dual -- the paper-paper model. In other settings, where higher quality reviewer-reviewer similarities are available, the relative merit of the model may increase.

\subsection{Putting it all together}
The overall model benefits from integrating into it the reviewer-reviewer component by the combined rule:

\begin{align} \label{model8}
 \hat{r}_{ui} = &\mu + b_u + b_i + p_u^Tq_i + \sum_c \sigma_{ic} \theta_{uc} w_c + \gamma \frac{\sum_{j \in {\rm R}(u)}s_{ij} r_{uj}}{\alpha+\sum_{j \in {\rm R}(u)}s_{ij}} \notag \\
 & + \phi \frac{\sum_{v \in {\rm R}(i)}s_{uv} r_{vi}}{\beta+\sum_{v \in {\rm R}(i)}s_{uv}}
\end{align}

All parameters are learnt simultaneously by minimizing the associated squared error on the train set.
This is our final prediction rule, which delivers an average test RMSE of {\bf 0.6015}. In the following section, we will show how filling up the unknown preferences using this model provides flexibility that enables deriving better paper assignments. 

\section{Optimizing Paper Assignment}
%Conference management research styles itself towards the terms and resources used by an individual conference or group of conferences featured in that research. However, there are a number of concepts common to most or all existing approaches. It is useful to develop and formalize a common terminology. In addition to the topic-based data and collected reviewer bids discussed above, some conferences also provide conflict of interest data, keyword relevance, and other data as depicted in Figure \ref{fig:data}. 
Our predicted preference matrix is now suitable for use with any of the 
optimization algorithms in Section~\ref{optimizations}. Denoting the output
of our preference modeling as the
affinity matrix $\indata$, the assignment problem can be formulated as motivated
in Taylor~\cite{taylor2006}:
 %below will indicate the general use of one or more of the input criteria, as desired for any given conference. The specific choice of affinity measure does not matter to the optimization formulation; Taylor's~\cite{taylor2006} criteria can be expressed in terms of input data matrix $\indata$ and the assignments matrix $\textbf{R}$, as:
%
%\donc{I think max of the trace is in orig paper... if we want to argmax we can prob eliminate the LHS of objective function and just argmax the double sum?}
%\donc{Fixed the subscript notation to match Yehuda's; don't believe Yehuda refers to matrices, only to individual elements, so for now I retain \indata syntax, subject to change.}
\begin{align}
\label{linprog}
 \argmax_{\textbf{R}} \quad &\text{trace}\left(\indata^T \textbf{R}\right) = \argmax_{\textbf{R}} \sum_u{\sum_j{\indata_{uj} \textbf{R}_{uj}}}, \\
 &\text{where} \qquad \textbf{R}_{uj} \in \left[0,1\right] \qquad \forall u,j, \notag \\
 &\text{and} \quad \sum_j{\textbf{R}_{uj}} \leq c_p, \qquad \forall u,\notag \\
 &\text{and} \quad \sum_u{\textbf{R}_{uj}} \leq c_r, \qquad \forall j.\notag
\end{align}
\noindent
Here $c_p$ represents the desired number of reviews per paper, and $c_r$ is the desired maximum reviews per reviewer. The third and fourth lines in the equation above represent the constraints on the number of assignments for individual papers and to individual reviewers, respectively. Then the expression $\text{trace}\left(\indata^T \textbf{R}\right)$ represents the global sum of affinity, or happiness of all reviewers across all assigned papers. In particular, by using the (binary) assignments matrix $\textbf{R}$ as a factor, only the affinities from $\indata$ for reviewer-paper combinations that exist in the final assignments $\textbf{R}$ are counted in the sum.

This integer programming problem \eqref{linprog} is reformulated into an easier-to-manage linear programming problem by a series of steps, using the node-edge adjacency matrix, where every row corresponds to a node in $\textbf{R}$, and every column represents an edge \cite{taylor2006}. This reformulation is a bit more complicated, but 
essentially maps the problem into the domain of linear programming and hence
solvable via methods such as Simplex or interior point programming.
In particular, as Taylor shows in~\cite{taylor2006},
because the reformulated constraint matrix is \emph{totally unimodular}, there exists at least one globally optimal solution (assignment set) with integral (and due to the constraints, Boolean) coefficients.

% \yehudac{Probably remove this paragraph.}
% One consequence of using integral values in the affinity matrix $\indata$ is that the resulting optimal solution is not necessarily unique. Taylor handles this by slightly perturbing the values (randomly), in order to impose a rank on equivalently valued nodes. While this does ensure a unique optimal solution, it appears likely
% that a more deterministic method of imposing this rank on like-valued nodes may prove more useful as a solution to the original problem.
% 
\section{Experimental Results}
%\donc{Still need the permission citation here. Is comment below enough, or should ICDM's size be further stressed?}
%The results in this paper are based on supporting data obtained from the International Conference on Data Mining, for the year 2007 (ICDM'07~\cite{icdm2007}). This data is given in matrix form; it includes reviewers 'bid' or preference data, as well as topical expertise of reviewers, and content classification for papers. The preference data consists of reviewer responses indicating their willingness (or lack of it) to review each paper. Since ICDM includes a relatively large number of submitted papers, most reviewers bid on a small subset of papers (and some do not return bids at all). As a result, the preference matrix is sparse; out of 529 papers and 203 reviewers (107,387 potential combinations), 6,267 bids were entered (less than 6 percent).
We have already demonstrated the ability of our modeling to better
capture reviewer-paper preferences. But do the improved models lead to better
assignments? In other words, does the assignment algorithm leverage the improved
modeling of preferences in ways that improve end-assignment quality? The key
distinction is between {\it preferences} versus {\it assignments}, an aspect
that has not been emphasized in prior recommender systems research.

We study these issues in the context of the IEEE ICDM'07 conference data as 
described earlier. Data from real conferences is quite rare to come by
(e.g., acknowledged also in~\cite{mccallum2007}) and in the future we hope 
that more datasets will become available to boost recommender systems
research in conference management.

The primary questions we seek to investigate are:
\begin{enumerate}
\item Do our preference models lead to improved topical relevance of assignments?
\item Do our preference models lead to higher quality assignments?
\end{enumerate}

%\narenc{Deflect criticism that we have only one dataset to use.}\donc{Cite difficulty due to double-blind confid. (esp.~\cite{mccallum2007})}

%\narenc{Give master list of all configurations that were tried out and
%how only some of them are described in detail for want of space.}

%Recall that our integrated preference
%model learned earlier can also predict ratings for potential 
%assignments for which no expressed preferences exist. To prevent overfitting, 
%we shrink these predictions slightly towards the mean rating; an identical technique is used by~\cite{koren2008}. We then used the original preference values as the primary baseline input data, and the predicted results as secondary data.
%First, we subtracted the per-reviewer mean from each predicted rating to find the \textbf{residual} rating for each potential assignment combination. We also calculated \textbf{normalized} ratings for each reviewer, so that the sum of each reviewers predicted ratings was 1. We combined each of these matrices with the original preferences 
%to form our final input matrix $\indata$ into Taylor's
%optimization algorithm. These two methods of preparing the input 
%are referred to as {\bf Resid} and {\bf Norm} methods in this section.

We use our preference model \eqref{model8} to predict ratings for potential assignments for which no expressed preferences exist. Before doing assignments using Taylor's model \eqref{linprog}, it is important to balance the rating scale of various reviewers. For example, some reviewers are very cooperative and tend to give mostly high ratings, while others are more cautious and give  medium to low ratings. Taylor's model may concentrate only on reviewers with high ratings, which is undesirable. Thus, we suggest two alternative per-reviewer normalization strategies: 
\begin{enumerate}
\item
Subtract the per-reviewer mean from each predicted rating to find the \textbf{residual} rating for each potential assignment combination. (Henceforth dubbed as {\bf Resid}.) 
\item
Calculate \textbf{normalized} ratings for each reviewer, so that the sum of each reviewer's predicted ratings is 1.  (Henceforth dubbed as {\bf Norm}.)
\end{enumerate}
Regardless of the chosen normalization scheme, we add the normalized
predicted rating to the original preferences; 
unknown values in the original preference matrix are considered 
to be the mean rating value (2.5) to place them between the 
`Ok' and `Low' ratings.
This forms our final input matrix $\indata$, which we feed into Taylor's optimization algorithm. 

\subsection{Topical relevance}
%I believe section 5.1 shows that our models can improve upon the new topical metric at the same time as maintaining Taylor's original affinity metric.  It is true that Taylor's algorithm has the advantage of using the entire set of known ratings, while the learning portion of the models uses only 90\%; despite this disadvantage, these models actually slightly improve results in the affinity metric.  I think the fact that our models perform better in both metrics when considering ALL the assignments is an important result, despite the unlevel playing field, possibly due to the additional content data used by our models.
To assess the topical relevance of the assignments, we evaluate them in
terms of the mappings
between papers/\hskip0ex reviewers and subject categories.
For every (paper,reviewer) assignment,
we compute the dot product of the category vector
of the paper with the category vector of the reviewer, and sum these
dot products over the assignments made.
Specifically paper-subject scores are recorded on a 2/1/0 scale (primary versus
secondary versus neither) and reviewer-subject scores are recorded on a 1/-1/0
scale (interest versus conflict versus neither). In our dataset here,
every paper has exactly one primary and one secondary category and hence
the dot product can yield a number between
-3 (reviewer has a conflict with both primary and secondary paper categories) 
and 3 (reviewer has interest in both paper categories).
%In addition to Taylor's original 'affinity' or rating sum metric, we can evaluate assignments based on topic. This is accomplished by computing the dot product of topical vectors for each paper and the reviewers to which they are assigned. The sum of these dot products provides a measure of topical relevance for the assignments.
%By computing the dot product of vectors $D^{exp}_{i}$ \& $R_{i}$ (i.e. scores for a reviewer across all papers) for each reviewer $i$, we obtain a measure of topical relevance for assignments made to that reviewer. Combining these dot product scores for all reviewers yields a topical sum for the given assignment set $R$ that can be used to compare against a different set of assignments. 
While other topical measures are certainly possible, the dot product method 
captures the relevance or `on-topicness' of assignments made to each reviewer. 
We used a 90\% training-10\% test set split to learn our Norm and Resid models, and calculated the mean of the predicted ratings for each reviewer-paper pair across 100 iterations.

\begin{figure}[!ht]
\centering
\includegraphics[width=3.3in]{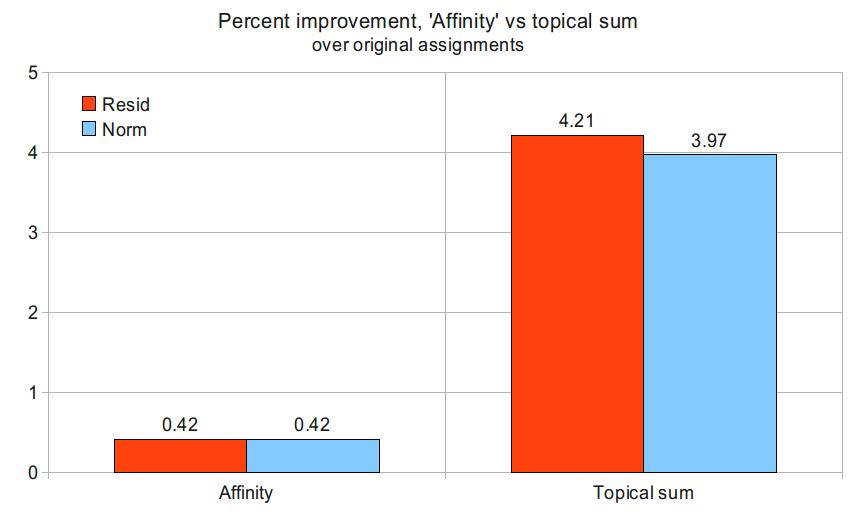}
\caption{Topical relevance of assignments made with our approach versus Taylor's original formulation.}
\label{fig:eval7}
\end{figure}

Fig.~\ref{fig:eval7} depicts the results in terms of percentage improvement
over the baseline Taylor approach (i.e., where only the original preferences
without any additional data were input to the LP). 
Note that the topical evaluation metric shows a measurable improvement using our 
modified ratings $\indata$ as input to Taylor's linear program.  Since our new models 
take topical relevance into account, this is not unexpected. However, we 
accomplished this topical optimization without degrading the Taylor algorithm's 
original `rating sum' objective; in fact, both the models considered here
slightly improve this objective as well (see Fig.~\ref{fig:eval7}).

%First, observe that we 
%see only a very small percent improvement over the traditional rating sum, i.e.,
%the paper-reviewer affinity, but a significant percent improvement in terms
%of topical relevance. These improvements are consistent over both our
%methods of preparing the input matrix.
%This demonstrates that our preference model not only makes
%better paper-reviewer predictions but that these predictions are effectively
%harnessed by the linear programming formulation to make topically relevant
%assignments.
%

\subsection{Assignment Quality}

The common train-test split methodology, which was used in Section \ref{sec:Prefmodels}, is also useful for assessing assignment quality.
Both prediction algorithm \eqref{model8} and assignment algorithm \eqref{linprog} cannot see the given preferences within the test set. Clearly, the elimination of the test set's preferences limits the flexibility of the assignment algorithm, as it has a lower number of favorable preferences from which to choose. However, the prediction model fills this gap by providing estimates to all missing preferences, including those in the test set. 
This simulates the real life scenario, where the given reviewer ratings (corresponding to the training set) are limiting the possibilities of assignment algorithm, but by revealing more ratings to the algorithms (including the test set) they gain the flexibility to provide better assignments.

As the proportion of the test set increases, we take away more available preferences, which simulates an increasingly harsher assignment environment. Accordingly, we evaluated several possible proportions, ranging from 50\% of the given preferences within the test set, to 30\% of preferences in the test set. In each experiment, we employed a series of 20 random train-test split and evaluated assignment quality. The baseline is Taylor's original algorithm, where all missing ratings, including those in the test set, are treated as ``unknowns.'' We compare this baseline against the two aforementioned alternatives, Resid and Norm.

\begin{figure}[!ht]
\centering
\includegraphics[width=3.5in]{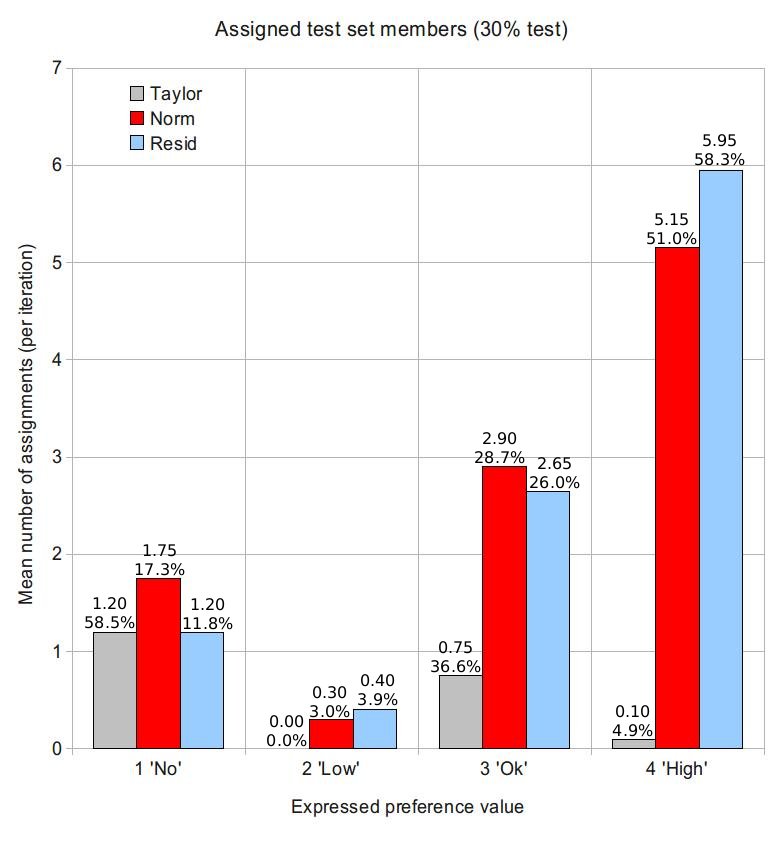}
\caption{Evaluating the assignments made by the unmodified Taylor algorithm and the
new preference models w.r.t. reviewers' four categories of
preferences, using a 70-30 test-training set split, averaged across 20 iterations. Mean assignments per iteration, and each value's percent of assignments for each iteration, are indicated above each bar.}
\label{fig:eval8_70-30}
\end{figure}

\begin{figure}[!ht]
\centering
\includegraphics[width=3.5in]{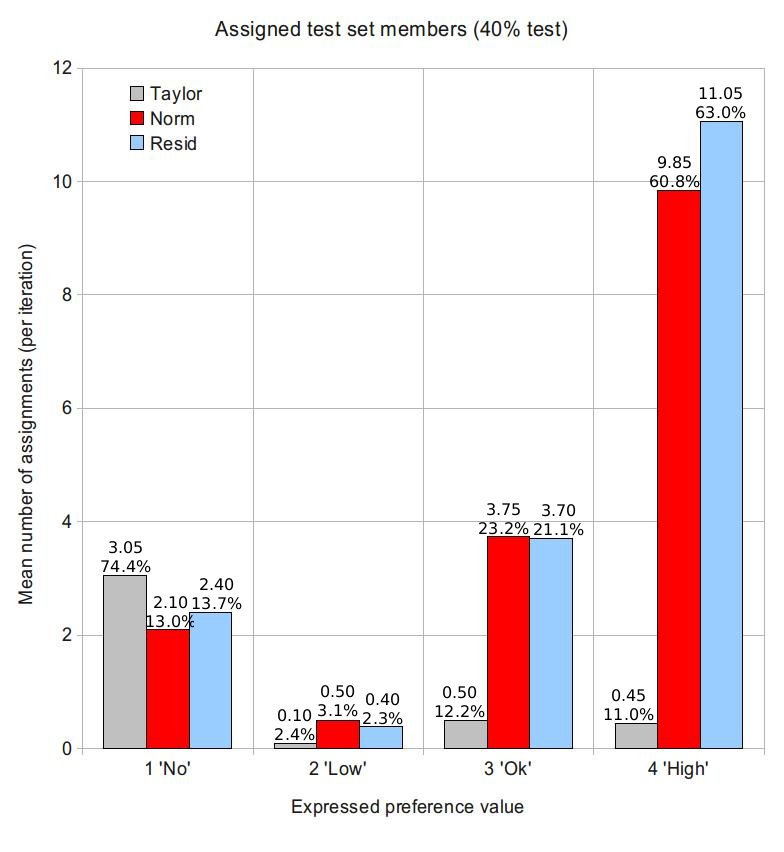}
\caption{Evaluating the assignments made by the unmodified Taylor algorithm and the
new preference models, using a 60-40 test-training set split.}
\label{fig:eval8_60-40}
\end{figure}

\begin{figure}[!ht]
\centering
\includegraphics[width=3.5in]{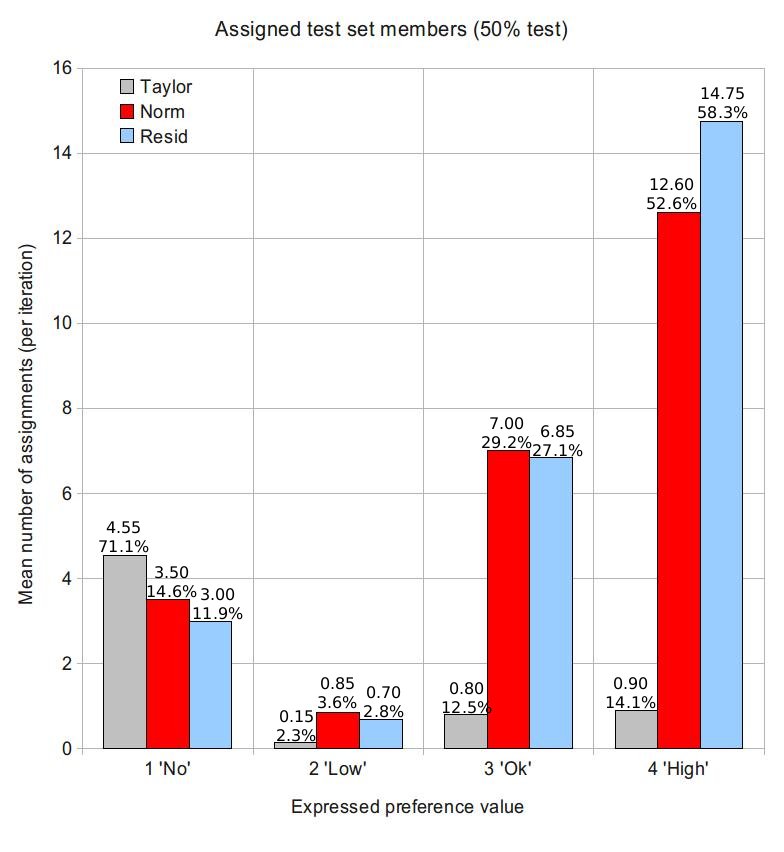}
\caption{Evaluating the assignments made by the unmodified Taylor algorithm and the
new preference models, using a 50-50 test-training set split.}
\label{fig:eval8_50-50}
\end{figure}

We evaluate quality of assignments by their ability to make good use of the hidden ratings in the test set.  The results are presented in 
%Figures~\ref{fig:eval8_50-50}, 
%\ref{fig:eval8_60-40}, \& 
%\ref{fig:eval8_70-30}, 
Figs.~\ref{fig:eval8_70-30}, 
\ref{fig:eval8_60-40}, \& 
\ref{fig:eval8_50-50}, 
and were fairly consistent over the different proportions of the test set. As illustrated here, the predominant number (around 60-70\%)  of test assignmentsmade using the original preference matrix fall in the unpreferred (``No'') category. On the other hand, when imputing the missing ratings, using either Resid or Norm, the balance completely changes in favor of higher quality preferences. Resid makes about 50-60\% of test assignments out of the highest quality ratings (``High''), and only about 15\% of test assignments are bad (``No''). Norm is close, but not quite as good as Resid, a difference that should be further investigated over additional datasets. Overall we find the results strongly support our goal to increase assignment quality by providing more flexibility with additional ratings from which to choose.

% \begin{table}
% \caption{Percentage of assignments made from the `unknown' test set members, broken down by actual known preference values. Our methods show improvement over the unmodified Taylor LP, assigning a much higher percentage of `preferred' papers.}
% \label{table:table2}
% \begin{center}
% \begin{tabular}{|l|cccc|} \hline
% \% assignments & 1 `No' & 2 `Low' & 3 `Ok' & 4 `High' \\
% from test set& & & &  \\ \hline
% Taylor & 59.91\% & 1.42\% & 30.19\% & 8.49\% \\
% Norm & 16.45\% & 4.09\% & 25.27\% & 54.19\% \\
% Resid & 11.15\% & 3.18\% & 24.63\% & 61.04\% \\ \hline
% \end{tabular}
% \end{center}
% \end{table}
%

\section{Discussion}
We have investigated the modeling of paper-reviewer preferences within a conference management system. The very limited data, typical to this context, requires identifying and exploiting multiple sources of information within a hybrid recommendation model. The proposed models provide improved predictions of reviewer preferences. More importantly, we showed how the improved modeling of such preferences can lead to improvements in actual review assignments. Encouraging experimental results demonstrate that the improved modeling
can be well worth the effort in ensuring satisfaction of conference program committee reviewers. A key question for future work is to provide theoretical justification for the empirical evidence presented here. We also intend to field the recommendation capabilities presented here in a real conference management system and gain further insights into the issues involved.

\section*{Acknowledgements}
The authors acknowledge the approval of Prof. Xindong Wu, Steering Committee
chair of the IEEE ICDM conference series for the use of preference/bid data collected
during the ICDM'07 conference reviewing process, and associated information
about papers/reviewers. All datasets were suitably
anonymized before the modeling and analysis steps conducted here.

\bibliographystyle{abbrv}
\bibliography{paper}

\begin{thebibliography}{10}

\bibitem{basu2001}
C.~Basu, H.~Hirsh, W.~Cohen, and C.~Nevill-Manning.
\newblock Technical paper recommendation: a study in combining multiple
  information sources.
\newblock {\em Journal of {AI} Research}, pages 231--252, 2001.

\bibitem{benferhat2001}
S.~Benferhat.
\newblock Conference paper assignment.
\newblock {\em International Journal of Intelligent Systems}, 16(10):1183,
  2001.

\bibitem{lda}
D.~Blei, A.~Ng, and M.~Jordan.
\newblock Latent {D}irichlet allocation.
\newblock {\em Journal of Machine Learning Research}, 3:993--1022, 2003.

\bibitem{canny}
J.~Canny.
\newblock Collaborative filtering with privacy via factor analysis.
\newblock In {\em Proc. {SIGIR{'}02}}, pages 238--245, 2002.

\bibitem{dumais1992}
S.~T. Dumais and J.~Nielsen.
\newblock Automating the assignment of submitted manuscripts to reviewers.
\newblock In {\em Proc. {SIGIR {'}92}}, pages 233--244, 1992.

\bibitem{tapestry}
D.~Goldberg, D.~Nichols, B.~Oki, and D.~Terry.
\newblock Using collaborative filtering to weave an information tapestry.
\newblock {\em Commun. of the ACM}, 35:61--70, 1992.

\bibitem{goldsmith2007}
J.~Goldsmith and R.~H. Sloan.
\newblock The {AI} conference paper assignment problem.
\newblock In {\em Pref. Handling for {AI}, Papers from the {AAAI} Workshop},
  2007.

\bibitem{hartvigsen1999}
D.~Hartvigsen, J.~C. Wei, and R.~Czuchlewski.
\newblock The conference paper-reviewer assignment problem.
\newblock {\em Decision Sciences}, 30(3):865--876, 1999.

\bibitem{plsa_cf}
T.~Hofmann.
\newblock Latent semantic models for collaborative filtering.
\newblock {\em {ACM} Transactions on Info. Systems}, 22:89--115, 2004.

\bibitem{karp1973}
J.~E. Hopcroft and R.~M. Karp.
\newblock An $n^{2.5}$ algorithm for maximum matching in bipartite graphs.
\newblock {\em {SIAM} Journal on Computing}, 18:225--231, 1973.

\bibitem{grouplens}
J.~A. Konstan, B.~N. Miller, D.~Maltz, J.~L. Herlocker, L.~R. Gordon, and
  J.~Riedl.
\newblock {G}roup{L}ens: applying collaborative filtering to usenet news.
\newblock {\em Commun. of the {ACM}}, 40(3):77--87, 1997.

\bibitem{koren2008}
Y.~Koren.
\newblock Factorization meets the neighborhood: a multifaceted collaborative
  filtering model.
\newblock In {\em Proc. {KDD{'}08}}, pages 426--434, 2008.

\bibitem{kuhn1955}
H.~W. Kuhn.
\newblock The {H}ungarian method for the assignment problem.
\newblock {\em Naval Research Logistics Quarterly}, 2:83--97, 1955.

\bibitem{GRAPE}
N.~D. Mauro, T.~M.~A. Basile, and S.~Ferilli.
\newblock {GRAPE}: an expert review assignment component for scientific
  conference management systems.
\newblock In {\em Proc. {IEA/AIE{'}2005}}, pages 789--798, 2005.

\bibitem{rmse-bad}
S.~McNee, J.~Riedl, and J.~Konstan.
\newblock {Being accurate is not enough: how accuracy metrics have hurt
  recommender systems}.
\newblock In {\em CHI Extended Abstracts}, pages 1097--11101, 2006.

\bibitem{mccallum2007}
D.~Mimno and A.~McCallum.
\newblock Expertise modeling for matching papers with reviewers.
\newblock In {\em Proc. {KDD{'}07}}, pages 500--509, 2007.

\bibitem{paterek}
A.~Paterek.
\newblock Improving regularized singular value decomposition for collaborative
  filtering.
\newblock In {\em Proc. {KDD} Cup and Workshop}, 2007.

\bibitem{popescul2001}
R.~Popescul, L.~H. Ungar, D.~M. Pennock, and S.~Lawrence.
\newblock Probabilistic models for unified collaborative and content-based
  recommendation in sparse-data environments.
\newblock In {\em Proc. of the 17th Conf. on Uncertainty in {AI}}, pages
  437--444, 2001.

\bibitem{icdm2007}
N.~Ramakrishnan, O.~Zaiaine, Y.~Shi, C.~Clifton, and X.~Wu.
\newblock Proc. {ICDM{'}07}, 2007.

\bibitem{rigaux2004}
P.~Rigaux.
\newblock An iterative rating method: application to web-based conference
  management.
\newblock In {\em Proc. {SAC{'}04}}, pages 1682--1687, 2004.

\bibitem{toronto}
R.~Salakhutdinov and Mnih.
\newblock Probabilistic matrix factorization.
\newblock In {\em Proc. {NIPS{'}07}}, pages 1257--1264, 2008.

\bibitem{rbm}
R.~Salakhutdinov, A.~Mnih, and G.~Hinton.
\newblock Restricted boltzmann machines for collaborative filtering.
\newblock In {\em Proc. 24th Annual Intl. Conf. on Machine Learning}, pages
  791--798, 2007.

\bibitem{item-item}
B.~Sarwar, G.~Karypis, J.~Konstan, and J.~Riedl.
\newblock Item-based collaborative filtering recommendation algorithms.
\newblock In {\em Proc. 10th Intl. Conf. on the World Wide Web}, pages
  285--295, 2001.

\bibitem{gravity-sigkddexplore}
G.~Takacs, I.~Pilaszy, B.~Nemeth, and D.~Tikk.
\newblock Major components of the gravity recommendation system.
\newblock {\em {SIGKDD} Explorations}, 9:80--84, 2007.

\bibitem{taylor2006}
C.~J. Taylor.
\newblock On the optimal assignment of conference papers to reviewers.
\newblock Technical Report MS-CIS-08-30, University of Pennsylvania, 2008.

\bibitem{weicroft2006}
X.~Wei and W.~B. Croft.
\newblock {LDA}-based document models for ad-hoc retrieval.
\newblock In {\em Proc. {SIGIR {'}06}}, pages 178--185, 2006.

\bibitem{yarowsky1999}
D.~Yarowsky and R.~Florian.
\newblock Taking the load off the conference chairs: towards a digital
  paper-routing assistant.
\newblock In {\em Proc. {EMNLP}'99.}, 1999.

\end{thebibliography}

\end{document}